\documentclass[useAMS,usenatbib]{mn2e}

\usepackage[]{natbib}
\usepackage{amsmath} 
\usepackage{amssymb} 
\usepackage{graphicx} 
\usepackage{indentfirst} 
\usepackage{pdflscape} 
\usepackage{placeins} 
\usepackage{nicefrac} 
\usepackage{afterpage} 
\usepackage{rotating} 
\usepackage{booktabs} 
\usepackage{rotating} 
\usepackage{color} 
\newcommand{\kms}{${\rm km\,s}^{-1}$} 
 
\newcommand{\lya}{\mbox{${\rm Ly}\alpha$}}

\title[]{On the possible environmental effect in distributing heavy elements beyond individual gaseous halos} 
\author[]{Sean D. Johnson$^{1}$\thanks{E-mail: seanjohnson@uchicago.edu}, Hsiao-Wen Chen$^{1}$, and John S. Mulchaey$^{2}$\\
$^{1}$Department of Astronomy \& Astrophysics and Kavli Institute for Cosmological Physics, The University of Chicago, Chicago, IL 60637, USA\\
$^{2}$The Observatories of the Carnegie Institution for Science, 813 Santa Barbara Street, Pasadena, CA 91101, USA}
\begin{document}
\date{\today}

\pagerange{
\pageref{firstpage}--
\pageref{lastpage}} \pubyear{2014}

\maketitle

\label{firstpage}
\begin{abstract}
We present a study of extended galaxy halo gas through
H\,I and O\,VI absorption over two decades in projected distance at $z\approx0.2$.
The study is based on a sample of $95$ galaxies from a highly complete ($>80\%$) survey of
faint galaxies ($L>0.1L_*$) with archival quasar absorption
spectra and $53$ galaxies
from the literature.
A clear anti-correlation is found between H\,I (O\,VI) column density
and virial radius normalized projected distance, $d/R_{\rm h}$.
Strong H\,I (O\,VI) absorption systems with column densities
greater than $10^{14.0}$ ($10^{13.5}$) cm$^{-2}$ are found
for $48$ of $54$ ($36$ of $42$) galaxies at $d<\,R_{\rm h}$
indicating a mean covering fraction of $\langle\kappa_{\rm H\,I}\rangle=0.89$
($\langle\kappa_{\rm O\,VI}\rangle=0.86$).
O\,VI absorbers are found at $d\approx R_{\rm h}$,
beyond the extent observed for lower ionization species.
At $d/R_{\rm h}=1-3$ strong H\,I (O\,VI) absorption
systems are found for only $7$ of $43$ ($5$ of $34$) galaxies
($\langle\kappa_{\rm H\,I}\rangle=0.16$
and $\langle\kappa_{\rm O\,VI}\rangle=0.15$).
Beyond $d=3\,R_{\rm h}$,
the H\,I and O\,VI covering fractions decrease to levels consistent with
coincidental systems.
The high completeness of the galaxy survey enables an investigation
of environmental dependence of extended gas properties.
Galaxies with nearby neighbors exhibit a modest increase in O\,VI
covering fraction at $d>R_{\rm h}$ compared to isolated galaxies ($\kappa_{\rm OVI}\approx0.13$ versus $0.04$) but no excess H\,I absorption.
These findings suggest that environmental
effects play a role in distributing heavy elements beyond the
enriched gaseous halos of individual galaxies.
Finally, we find that differential H\,I and O\,VI absorption
between early- and late-type galaxies
continues from $d<R_{\rm h}$ to $d\approx3\,R_{\rm h}$.

\end{abstract}

\section{Introduction}
\label{section:introduction}

\begin{table*}
\centering
\caption{Summary of galaxy and absorber properties.}
\label{table:galaxies}
\centering \resizebox{6.9in}{!}{
\begin{tabular}{lccccccccccc}
\hline
& \multicolumn{2}{c}{Galaxy}  \\ \cline{2-3}
Quasar & R.A. & Dec. & & & & & & $d$ & & \multicolumn{2}{c}{$\log\,N/{\rm cm}^{-2}$} \\ \cline{11-12}
sightline & (J2000) & (J2000) & $z_{\rm gal}$ & $\log\,(M_*/M_\odot)_{gr}$ & class$^{\rm a}$ & envi.$^{\rm b}$ & survey & (kpc) & $d/R_{\rm h}$ & H\,I$^{\rm c}$ & O\,VI$^{\rm c}$ \\
\hline \hline
J0042$-$1037     & 00:42:22.27  & $-$10:37:35.2   &  $0.0950$ & $9.5$ & Late & I & COS-Halos &  $15$ & $0.1$ & $14.80-18.50$ & $14.70 \pm 0.22$\\
J0226+0015     & 02:26:12.98  & +00:15:29.1   &  $0.2274$ & $10.4$ &Early & I & COS-Halos &  $81$ & $0.4$ & $14.36 \pm 0.06$ & $<13.12$\\
HE0226$-$4110    & 02:27:46.27  & $-$40:53:15.1   &  $0.1120$ & $10.4$ &Early & I & IMACS &  $826$ & $4.1$ & $<12.88$ & $<12.90$\\
HE0226$-$4110    & 02:27:57.47  & $-$40:57:25.7   &  $0.2952$ & $8.6$ & Late & I & IMACS &  $884$ & $9.8$ & $13.36 \pm 0.05$ & $<12.75$\\
HE0226$-$4110    & 02:28:01.44  & $-$40:57:55.7   &  $0.3863$ & $9.8$ & Late & I & IMACS &  $846$ & $6.5$ & $13.39 \pm 0.04$ & $<12.66$\\
HE0226$-$4110    & 02:28:04.31  & $-$40:55:57.4   &  $0.1772$ & $10.5$ &Early & NI & IMACS &  $435$ & $2.1$ & $<12.23$ & $<13.07$\\
HE0226$-$4110    & 02:28:04.52  & $-$40:57:02.3   &  $0.3971$ & $11.1$ & Late & NI & IMACS &  $648$ & $1.8$ & $13.99 \pm 0.02$ & $13.57 \pm 0.09$\\
HE0226$-$4110    & 02:28:09.37  & $-$40:57:44.5   &  $0.3341$ & $9.7$ & Late & NI & IMACS &  $346$ & $2.6$ & $<12.36$ & $<12.75$\\
HE0226$-$4110    & 02:28:10.23  & $-$40:55:48.1   &  $0.1248$ & $8.5$ & Late & I & IMACS &  $230$ & $2.3$ & $12.88 \pm 0.05$ & $<12.84$\\
HE0226$-$4110    & 02:28:10.50  & $-$40:58:57.3   &  $0.3268$ & $9.4$ &Early & I & IMACS &  $547$ & $4.7$ & $<12.35$ & $<13.05$\\
HE0226$-$4110    & 02:28:11.32  & $-$40:56:36.7   &  $0.2804$ & $8.4$ & Late & I & IMACS &  $244$ & $2.8$ & $13.73 \pm 0.01$ & -1\\
HE0226$-$4110    & 02:28:11.59  & $-$40:58:53.8   &  $0.2492$ & $8.6$ & Late & I & IMACS &  $419$ & $4.5$ & $13.38 \pm 0.03$ & $<13.16$\\
HE0226$-$4110    & 02:28:13.89  & $-$40:59:46.2   &  $0.3939$ & $10.2$ & Late & I & IMACS &  $812$ & $5.2$ & $<12.36$ & $<12.62$\\
HE0226$-$4110    & 02:28:14.56  & $-$40:57:22.7   &  $0.2065$ & $8.9$ & Late & NI & IMACS &  $37$ & $0.4$ & $15.27 \pm 0.06$ & $14.37 \pm 0.01$\\
HE0226$-$4110    & 02:28:16.29  & $-$40:57:27.2   &  $0.2678$ & $10.3$ &Early & I & IMACS &  $75$ & $0.4$ & $<12.97$ & $<12.87$\\		
\hline \multicolumn{12}{l}{\bf Notes} \\
		\multicolumn{12}{l}{The full table is available in the online version of the paper.} \\
		\multicolumn{12}{l}{$^\mathrm{a}$ Galaxy classification with ``Early'' for early-type, absorption-line dominated galaxies and ``Late'' for late-type, emission-line dominated galaxies} \\
		\multicolumn{12}{l}{$^\mathrm{b}$ Galaxy environment class as defined in the text with ``I'' for isolated, ``NI'' for non-isolated, and ``A'' for ambiguous cases.} \\
		\multicolumn{12}{l}{$^\mathrm{c}$ Total column density measured in the COS sightline within $\Delta v = \pm 300$ \kms\, of the galaxy systemic redshift. In cases where O\,VI or H\,I column} \\
		\multicolumn{12}{l}{\,\,\,\, densities cannot be measured due to contamination, the value is set to -1.} \\
\end{tabular}
}
\end{table*}

Galaxies grow by a combination of gas accretion from
the intergalactic medium (IGM) and mergers.
After heavy element enrichment, it is expected that gas is subsequently
ejected back to the IGM through starburst and AGN driven
winds \citep[e.g.][]{Murray:2011}
or by stripping during galaxy interactions
\citep[e.g.][]{Wang:1993, Gauthier:2013}.
The low-density gas of the circum-galactic medium (CGM)
is at the nexus of these baryon cycles,
and it represents a potentially significant, 
multiphase reservoir
that can fuel star-formation on long time-scales \citep[e.g.][]{Maller:2004}.
Moreover, the CGM constitutes a sensitive laboratory for studying
the physical processes that control the inflow and outflow of baryons from galaxies
and for testing sub-grid feedback recipes implemented
in simulations of galaxy
evolution \citep[e.g.][]{Hummels:2013, Ford:2013,Cen:2013,Shen:2013,Agertz:2014}.

The density of the CGM is nearly always too
low to be studied in emission with existing facilities.
Nevertheless, a great deal of progress has
been made via UV absorption-line spectroscopy
of background objects at low projected
distances, $d$, from foreground galaxies.
Observations of the H\,I Lyman series
\citep[e.g.][]{Chen:1998, Tripp:1998, Wakker:2009,
Stocke:2013,
Rudie:2013, Tumlinson:2013},
the Mg\,II doublet
\citep[e.g.][]{Bowen:1995, Chen:2010a, Gauthier:2010, Bordoloi:2011},
the C\,IV doublet
\citep[e.g.][]{Chen:2001a, Borthakur:2013, Liang:2014, Bordoloi:2014},
and the O\,VI doublet \citep[e.g.][]{Chen:2009, Wakker:2009, Prochaska:2011,
Tumlinson:2011, Mathes:2014, Stocke:2014, Turner:2014} have been particularly fruitful.

Cool, metal-enriched gas traced by Mg\,II, C\,II, Si\,II, Si\,III, and C\,IV
absorption is observed out to impact
parameters of $0.7$ times the galaxy virial radius ($d\approx0.7\,R_{\rm h}$) around both late-type
and early-type galaxies of $L\lesssim L_*$ alike\footnote{There is evidence that C\,IV absorption is more extended around
starburst galaxies \citep{Borthakur:2013}, and more massive luminous red galaxies exhibit reduced Mg\,II absorption \citep{Gauthier:2010}.}.
At larger projected distances, the incidence of these absorption
species decreases sharply to near zero.
\cite{Liang:2014} present two possible explanations for
this ``metal-boundary'': 
(1) an inability of galactic outflows and satellite accretion to efficiently
deposit heavy element enriched gas at distances greater than $0.7\,R_{\rm h}$
or (2) an inability of cool-warm gas clouds to form and survive at larger distances.

In contrast, highly-ionized and enriched gas traced by the O\,VI doublet
is found to reflect the bimodal distribution of galaxies
in color-magnitude diagrams. In particular, strong O\,VI systems are found
at high incidence out to $d=150$ kpc around star-forming
galaxies but at lower incidence around more quiescent ones \citep{Chen:2009, Tumlinson:2011}.
Moreover, O\,VI detections have been reported around sub-$L_*$ galaxies
out to $d=300$ kpc which corresponds to $1-3\,R_{\rm h}$ \citep{Prochaska:2011}.
However, the sample of galaxies probed at such large distances by absorption
spectroscopy is small and the spatial extent of O\,VI bearing gas around galaxies
remains an open question.

To examine the spatial dependence of O\,VI absorption around galaxies,
we have assembled a sample
of $95$ galaxies that are probed in absorption by quasar sightlines with
UV spectra from the Cosmic Origins Spectrograph \citep[COS;][]{Green:2012} on the
{\it Hubble Space Telescope} (HST).
These galaxies are drawn from a highly complete
($\gtrsim80\%$ complete for galaxies of $L>0.1L_*$ at $z<0.4$ and $d<500$ kpc)
survey of faint galaxies in the fields of four
high signal-to-noise COS quasar sightlines.
The high completeness of the galaxy survey enables
a detailed investigation of possible environmental
dependence of extended gas around galaxies.
To increase the number of massive galaxies in the study,
we include eleven galaxies with stellar masses of $\log\,M_*/M_\odot>11$
from the Sloan Digital Sky Survey \citep[SDSS;][]{York:2000}
probed in absorption with archival COS quasar spectra.
We refer to these galaxy samples collectively
as the extended-CGM (eCGM) survey
and combine it with $42$ galaxies from the COS-Halos program
\citep[][]{Werk:2012, Tumlinson:2013}
which are primarily at $d<150$ kpc.
Together, the two samples contain $148$ galaxies
and enable an investigation of
the absorption properties of galaxy halos
over two decades in projected distance.

The paper proceeds as follows:
In Section \ref{section:galaxies} we describe the galaxy sample.
In Section \ref{section:absorption} we describe the corresponding
absorption-line measurements.
In Section \ref{section:extent} we characterize the observed
H\,I and O\,VI absorption as a function of projected distance.
In Section \ref{section:discussion}
we discuss the implications of our survey results.
Throughout the paper, we adopt a $\Lambda$ cosmology with $\Omega_{\rm m}=0.3$, $\Omega_{\Lambda} = 0.7$, and $H_0=70$ \kms\,${\rm Mpc}^{-1}$.

\section{The extended-CGM galaxy sample}
\label{section:galaxies}

To assemble the eCGM galaxy sample,
we combined our own absorption-blind survey data targeting galaxies
of $r_{\rm AB}<23$ mag in the fields
of HE\,$0226$-$4110$, PKS\,$0405$-$123$,
LBQS\,$1435$-$0134$, and PG\,$1522$+$101$
with spectroscopic galaxies in the Sloan Digital Sky Survey Data Release 10
\citep[SDSS;][]{York:2000, Ahn:2014}.
The HE\,$0226$-$4110$, PKS\,$0405$-$123$,
LBQS\,$1435$-$0134$, and PG\,$1522$+$101$ fields
were selected because of the high completeness levels ($\gtrsim80\%$)
achieved by our surveys
for galaxies as faint as $L=0.1\,L_*$ at $z<0.4$ and $d<500$ kpc.
At smaller projected distances of $d=100$ and $200$ kpc, the survey
completeness increases to $100\%$ and $90\%$ respectively
 \citep[see Figure 2 of][]{Johnson:2013}\footnote{
Here, we adopt a characteristic luminosity of $L_*$ galaxies
of $M_r=-21.5$ based on the luminosity
function measurements from \cite{Dorta:2009} and \cite{Loveday:2012}.}.
We included SDSS galaxies to increase the number of eCGM
sample members of $\log\,M_*/M_\odot>11$.
The resulting galaxy sample is summarized in Table \ref{table:galaxies},
and its construction is described in this section.

Our galaxy surveys in the 
HE\,$0226$-$4110$, PKS\,$0405$-$123$,
LBQS\,$1435$-$0134$, and PG\,$1522$+$101$ fields were carried out
with the IMACS \citep{Dressler:2011} and LDSS3 spectrographs on
the Magellan Telescopes \citep{Shectman:2003} targeting
galaxies as far as $\Delta\theta=10'$ from the quasar sightline
(corresponding to $d=1.1$ Mpc at $z=0.1$).
The resulting redshift catalogs and corresponding
absorption-line measurements for the HE\,$0226$-$4110$, PKS\,$0405$-$123$
fields were published in
\cite{Chen:2009} and \cite{Johnson:2013}.
The published survey data include deep $B$-, $V$-, and $R$-band
photometry in the HE\,$0226$-$4110$ field.
The LBQS\,$1435$-$0134$, and PG\,$1522$+$101$
galaxy redshift catalogs and corresponding
IMACS $g$-, $r$-, and $i$-band photometry
will be published in Johnson et al. in prep.
For the PKS\,$0405$-$123$ field, we retrieved
$g$-, $r$-, and $i$-band images from the Mosaic II CCD
Imager on the Blanco $4$-m telescope from the NOAO archive
(PI: Brian Keeney; PID=2008B-0194).
The Mosaic II images consist of $7\times75$ second exposures in each bandpass.
We measured the galaxy photometry using Source Extractor \citep{Bertin:1996}
with settings described in \cite{Johnson:2014}.

To increase the number of massive galaxies in the eCGM sample,
we cross-matched galaxies of $\log\,M_*/M_\odot>11$ in the SDSS DR10 spectroscopic survey
with quasar sightlines that have public, high signal-to-noise
COS spectra published in \cite{Danforth:2014}. 
For the SDSS galaxies, we adopted $u$-, $g$-, $r$-, $i$-, and $z$-band
cmodel magnitudes\footnote{The results presented in this paper do not change if model rather than cmodel magnitudes are adopted for SDSS galaxies.}.

\begin{figure}
	\centering
	\includegraphics[scale=0.42]{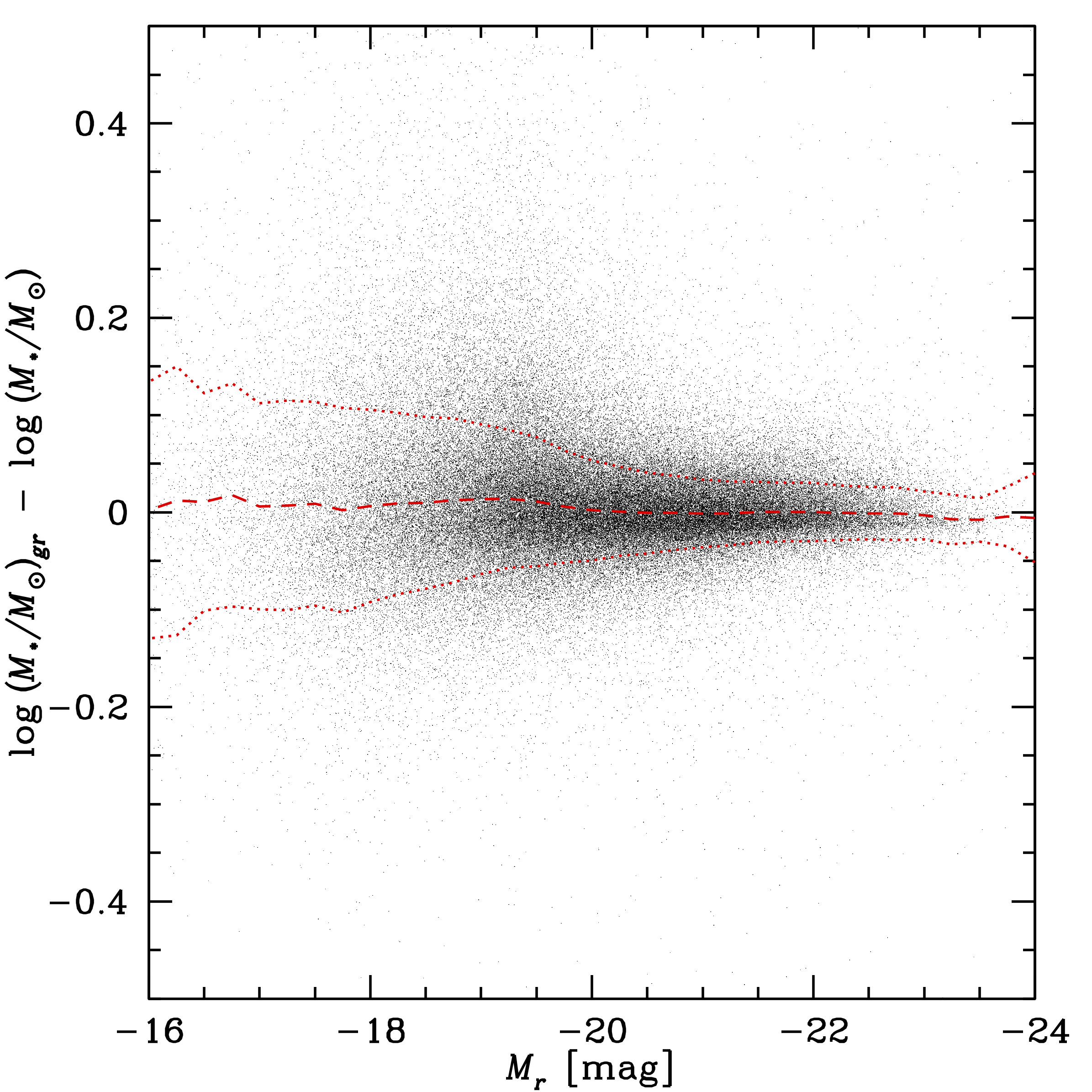}
	\caption{Residuals between rest-frame $g$- and $r$-band based
	              stellar masses estimated using
	              Equations \ref{equation:logML} and \ref{equation:logMstar}
	              and those from the NASA-Sloan atlas
	              versus absolute $r$-band magnitude, $M_r$.
	              Each point represents a NASA-Sloan galaxy.
	              The dashed red line shows the median residuals
	              in $M_r$ bins with widths of $0.25$ mags.
	              The dotted red lines mark $\pm1\sigma$ from the median in each bin.
	              Within the NASA-Sloan atlas, the systematic error is less than $0.02$ dex with
                      $1\sigma$ scatter of less than $0.15$ dex.} 
	\label{figure:logMstar}
\end{figure}

From this parent sample of galaxies,
we selected those with projected distances
from the quasar sightline $d<1$ Mpc, within the
redshift range $z\approx0.1-0.4$, and with redshift
at least $10,000$ \kms\, below the quasar redshift.
O\,VI detections have been reported at projected
distances of up to $300$ kpc from sub-$L_*$ galaxies
\citep{Prochaska:2011} which
corresponds to $2-3$ times the galaxy halo virial radius.
The $d=1$ Mpc cut was chosen to be sufficiently large to search
for absorption at $d/R_{\rm h}=2-3$ around more massive
galaxies that have virial radii of $R_{\rm h} \approx 300$ kpc.
The redshift lower limit was chosen so that the
O\,VI doublet falls within the COS wavelength range.
We note that the redshift range for which 
the O\,VI doublet is observable with COS
depends on the settings used for the observations.
The redshift lower limit therefore varied slightly from sightline to sightline.
The redshift upper limit was motivated by the depth
of our IMACS redshift surveys which
are sensitive to galaxies of $L>0.1\,L_*$  at $z\lesssim0.4$.
The requirement that the galaxy redshift be
at least $10,000$ \kms\, below the quasar redshift
was imposed to avoid confusion
with gas outflowing from the quasar \citep[e.g.][]{Wild:2008}.
In total, $106$ galaxies met these criteria with eleven
galaxies from the SDSS and $95$ from our
survey data.

For each galaxy, we measured the rest-frame,
$g$- and $r$-band absolute magnitudes
using the multi-band optical photometry\footnote{
Corrected for Milky Way extinction from \cite{Schlegel:1998}.}
described above and the kcorrect tool
\citep{Blanton:2007}.
We then estimated mass-to-light ratios
using the relation shown in
Equation \ref{equation:logML} which
is a fit of the stellar mass to $r$-band light ratio (in solar units)
for low redshift galaxies in the NASA-Sloan atlas \citep[e.g.][]{Maller:2009}.
For galaxies redder than $M_g-M_r=0.15$ in the rest-frame,
the fit is a quadratic in $M_g - M_r$.
For galaxies bluer than $M_g - M_r=0.15$, 
we set the mass-to-light ratio to $\log\,(M_*/L_r)=-0.6$
which corresponds to an imposed minimum age of $\approx0.3$ Gyr for a
simple stellar population with solar metallicity.

\begin{equation}
\label{equation:logML}
\resizebox{3.0in}{!}{$\displaystyle
\log\,(M_*/L_r) = \begin{cases} -0.6 & M_g - M_r<0.15 \\ 
-1.01 + 2.95(M_g - M_r)\dots  & M_g - M_r\geq0.15 \\
  - 1.67(M_g - M_r)^2 \end{cases}$}
\end{equation}
We then estimated the stellar mass of the eCGM galaxies using
Equation \ref{equation:logMstar} and the
mass-to-light ratios from Equation \ref{equation:logML}.
\begin{equation}
\label{equation:logMstar}
\log\,(M_*/M_\odot)_{gr} = 1.872 - 0.4M_r + \log\,(M_*/L_r)
\end{equation}

The rest-frame $g$- and $r$- band stellar mass estimates
from Equations \ref{equation:logML} and \ref{equation:logMstar}
reproduce the NASA-Sloan stellar masses which are
calculated with kcorrect, a \cite{Chabrier:2003} initial mass function
and GALEX $FUV+NUV$ and SDSS $ugriz$ photometry with
a systematic error of less than $0.02$ dex and $1\sigma$ scatter
of less than $0.15$ dex for galaxies of $M_r=-16$ to $-24$
as shown in Figure \ref{figure:logMstar}.
The NASA-Sloan atlas contains low-redshift galaxies of $z<0.055$
so the use of a mass-to-light ratio relation based on these galaxies
for the $z=0.1-0.4$ eCGM sample may result in systematic errors
due to galaxy evolution over cosmic time.
To estimate the magnitude of this systematic error, we compared
stellar masses calculated using 
Equations \ref{equation:logML} and \ref{equation:logMstar}
to those based on precise $ugriz$ photometry for $z=0.1-0.4$
galaxies  from the  PRIMUS galaxy survey \citep{Coil:2011}.
The systematic error introduced by using a $z<0.055$
relation for $z=0.1-0.4$ galaxies is less than $0.1$ dex over the
luminosity range spanned by the eCGM sample.

\begin{figure}
	\centering
	\includegraphics[scale=0.42]{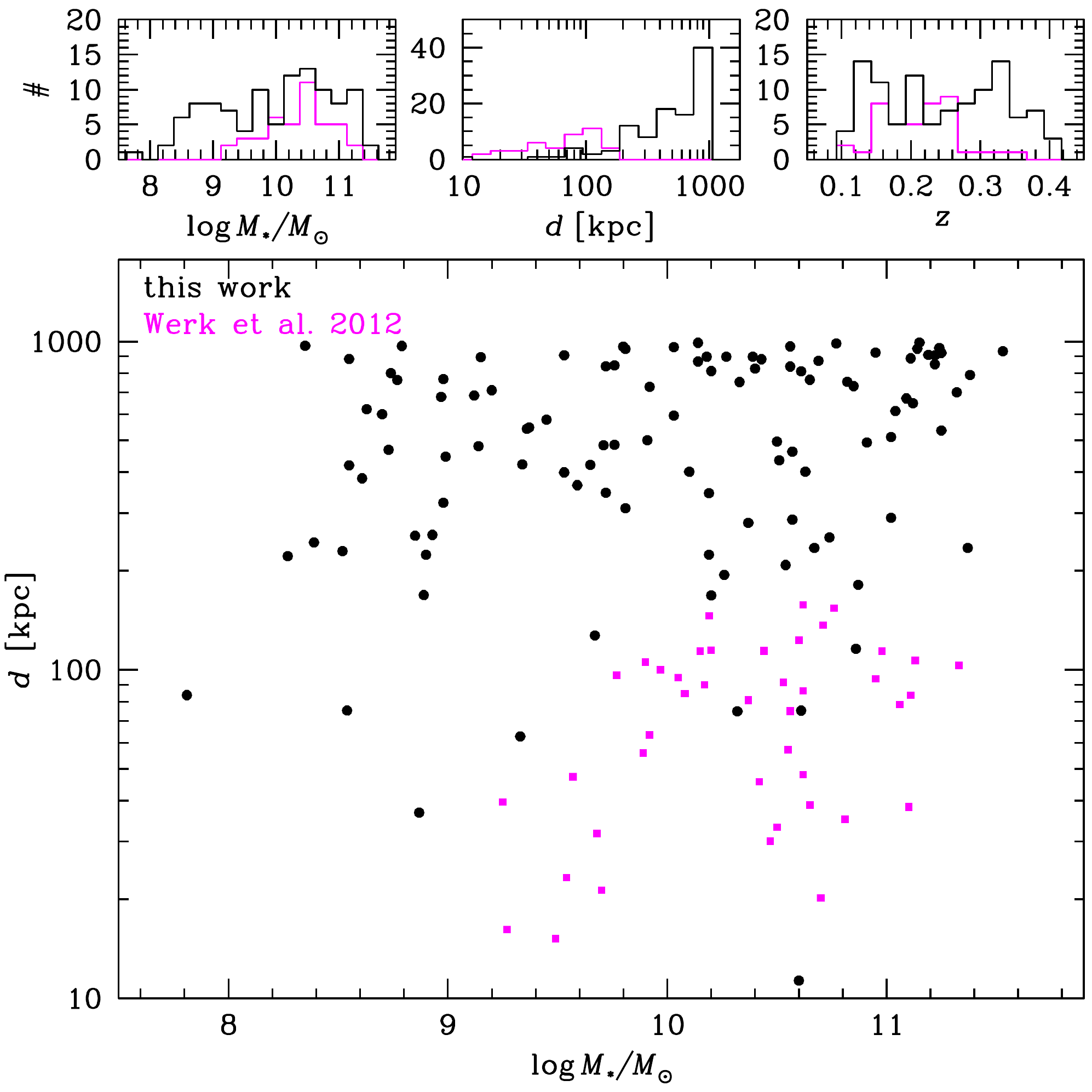}
	\caption{ Distribution of projected distance, $d$, between each galaxy
	               and the nearby COS quasar sightline versus stellar mass for
	               the eCGM galaxy sample (black circles) 
	               and the COS-Halos survey (magenta squares). 
	               The {\it top} three panels show the stellar mass,
	               projected distance, and redshift distributions of
	               the eCGM and COS-Halos galaxy samples.} 
	\label{figure:galaxies}
\end{figure}

The eCGM galaxy sample spans a stellar mass
range of $\log\,M_*/M_\odot=8.4-11.5$
with a median of $\log\,M_*/M_\odot=10.3$
as shown in Figure \ref{figure:galaxies}.

In order to compare absorption properties of galaxies
across a range of stellar mass, we measured projected distance
in units of the galaxy dark matter halo virial radius defined by \cite{Bryan:1998}
using the stellar-to-halo mass relation
from \cite{Kravtsov:2014}.
To ensure consistency across the sample, we
re-measured the virial radii of the COS-Halos galaxies
using the same methods.
We note that our virial radius estimates are as much as 
$50\%$ smaller than those of \cite{Tumlinson:2013}
because of the use of a different stellar-to-halo mass relation.
The differences between the stellar-to-hallo mass relations from
\cite{Kravtsov:2014} and previous work such as \cite{Moster:2010}
and \cite{Behroozi:2013} are driven by systematic errors in the
galaxy photometry in the catalogs used by \cite{Moster:2010}
and \cite{Behroozi:2013} \citep[see ][]{Bernardi:2013}.
The stellar-to-halo mass relation from \cite{Kravtsov:2014} is
in good agreement with independent constraints from weak
lensing \citep[e.g.][]{Reyes:2012}.

To search for correlations between CGM properties and
the presence of star-formation, we spectroscopically classified galaxies as
late-type (emission-line dominated) and early-type (absorption-line dominated)
based on a PCA fitting technique described in \cite{Chen:2009}.

Unlike some previous surveys,
we did not restrict the eCGM galaxy sample to isolated galaxies only.
Instead, we classified galaxies as ``non-isolated'' or ``isolated''
based on the presence or absence of nearby neighbors.
In the HE\,$0226$-$4110$, PKS\,$0405$-$123$,
LBQS\,$1435$-$0134$, and PG\,$1522$+$101$  fields,
where our galaxy survey data are highly complete,
we classified galaxies as ``non-isolated'' if they have a
spectroscopic neighbor within a projected distance of less than $500$ kpc,
a radial velocity difference of $|\Delta v| < 300$ \kms,
and with stellar mass of at least one-third of that of the eCGG or COS-Halos survey member.
Otherwise, we classified galaxies without such nearby neighbors as ``isolated''.

Among non-isolated galaxies, the relevant  projected distance
between the quasar sightline and the galaxy is not clear apriori,
and care must be taken in defining the distance to a non-isolated
galaxy to prevent a bias when comparing with isolated ones.
For example, a mass or light-weighted projected distance could be
large even though the gas detected in absorption may be more closely
associated with a less massive ``group'' member close to the quasar sightline.
{\it To avoid a bias that could make the gas around non-isolated galaxies
appear artificially extended, we define the distance to non-isolated galaxies
as the minimum value of $d/R_{\rm h}$ among ``group'' members}. This
distance definition is a self-consistent choice for comparing the spatial
extent of absorbing gas around isolated and non-isolated galaxies.

The available galaxy survey data
in the COS-Halos and SDSS fields are not sufficiently complete to
confidently classify galaxies by environment based
on spectroscopy alone. To classify these galaxies,
we performed a literature
search to identify members of previously known
galaxy groups \citep[e.g.][]{McConnachie:2009}
and clusters \citep[e.g.][]{Koester:2007}.
In addition, we visually inspected the SDSS
images in the vicinity of each galaxy to search
for likely neighbors based on galaxy
brightness, size, color, and SDSS photometric redshift. We classified galaxies
with such neighbors as non-isolated
and those without as isolated.

\begin{figure*}
	\centering
	\includegraphics[scale=0.65]{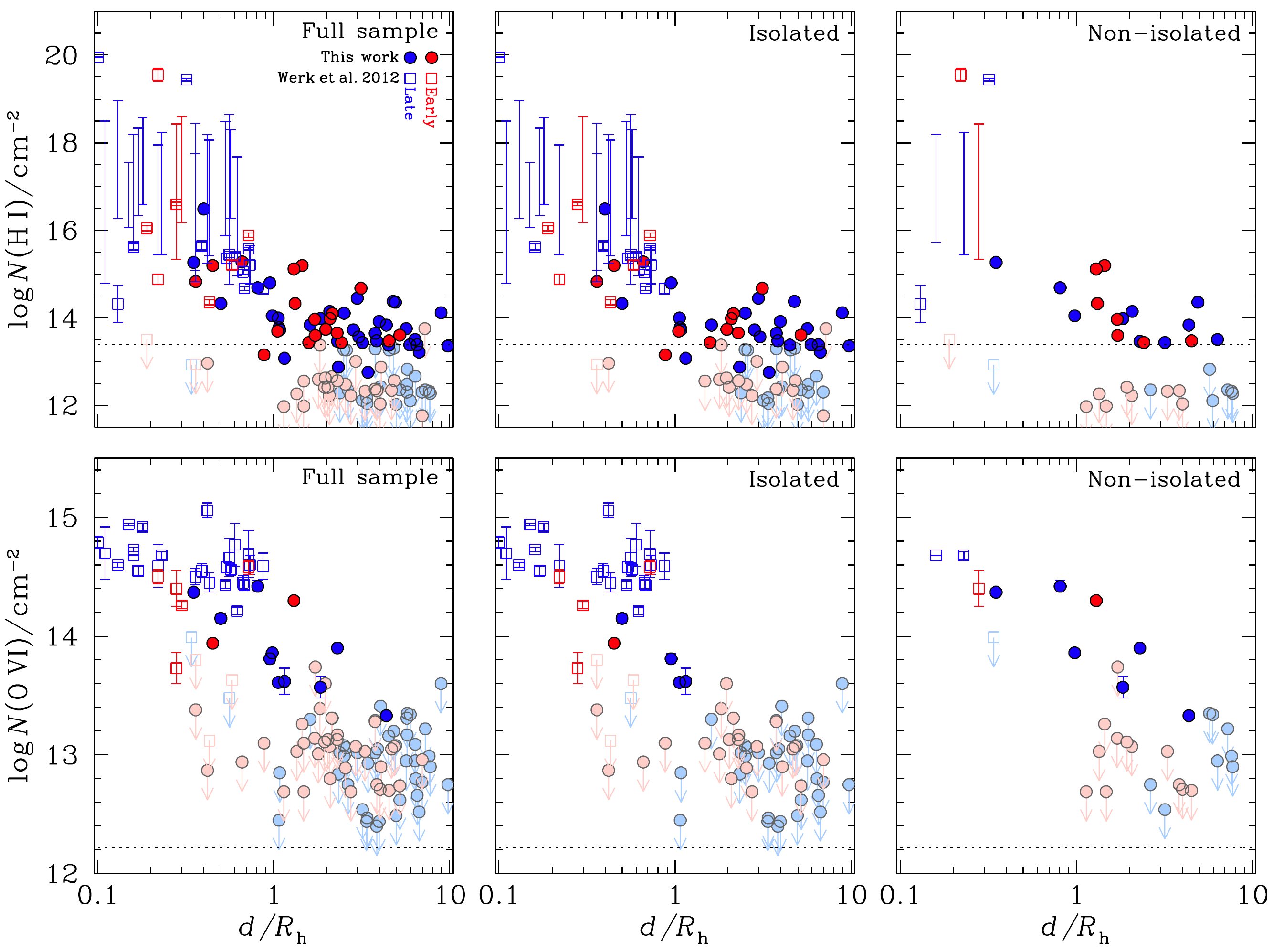}
	\caption{Column density within $\Delta v = \pm 300$ \kms\,
	versus projected distance measured in
	units of the galaxy virial radius with H\,I in the {\it top} panels
	and O\,VI on the {\it bottom}.
	The full galaxy samples are shown in the {\it left}
	column and are separated into isolated and non-isolated members in the {\it middle} and {\it right}
	columns respectively.
	Late- and early-type galaxies are shown in blue and red respectively.
	eCGM galaxies are displayed as filled circles and COS-Halos galaxies as open squares.
	Non-detections are shown as $2\sigma$ upper limits
	marked by downward arrows and lighter coloring. 
	The vertical error bars represent $1$$\sigma$ statistical uncertainties reported
	by VPFIT and do not include uncertainty due to continuum placement.
	Galaxies with upper limits greater than $\log N/{\rm cm}^{-2} = 14.0$
	are not considered to have sensitive constraints on the gas column density
	and are not displayed.
	The mean absorption expected from random sightlines based on the 
	$\frac{d^2\mathcal{N}}{dNdz}$ measurement from \protect \cite{Danforth:2014}
	is shown as horizontal
	dotted line.}
	\label{figure:d_Rh}
\end{figure*}

\section{Absorption data}
\label{section:absorption}
For each galaxy in the eCGM sample,
we searched for H\,I and O\,VI absorption in the COS quasar spectra
within a radial velocity interval of
$\Delta v = \pm 300$ \kms\, of the galaxy systemic
redshift. \cite{Tumlinson:2011} found that the radial velocity
distribution of O\,VI absorption-components
around galaxies at $d<150$ kpc is characterized by a
standard deviation of $\sigma\approx100$ \kms.
The $\Delta v = \pm 300$ \kms\, search window
was chosen to include the vast majority of absorption associated
with the eCGM and COS-Halos galaxies\footnote{The results presented in this paper do
not change if the velocity window is increased to $\Delta v = \pm 600$ \kms.}.
If H\,I or O\,VI absorption was detected, we adopted available
Voigt profile fitting results from \cite{Johnson:2013}, \cite{Tumlinson:2013}, \cite{Werk:2013}, and \cite{Savage:2014}.
When these measurements are not available, we performed our own
Voigt profile fitting using the VPFIT package \citep{Carswell:1987}
as described in \cite{Johnson:2013}.
If H\,I or O\,VI absorption was not detected,
we placed $2\sigma$ upper limits on the equivalent width,
integrating over a velocity interval of $75$ \kms\, for H\,I and $90$ \kms\, for O\,VI. 
We then converted this equivalent width limit to a column density
assuming that the gas is optically thin.
The $75$ \kms\, and $90$ \kms\, integration windows
correspond to the median full-width-at-half-maximum
of unsaturated H\,I and O\,VI absorption
components found in the eCGM survey.

For $19$ galaxies, \cite{Tumlinson:2013} report only lower limits on H\,I column
density due to saturation in all available Lyman series transitions.
For these galaxies, we place {\it upper limits} on the possible H\,I column
density based on the lack of damping wings detected in \lya.
To do so, we measured the \lya\, absorption equivalent width
and converted this to a H\,I column density upper limit
assuming a single component with doppler width $b=20$ \kms.

\begin{figure*}
	\centering
	\includegraphics[scale=0.65]{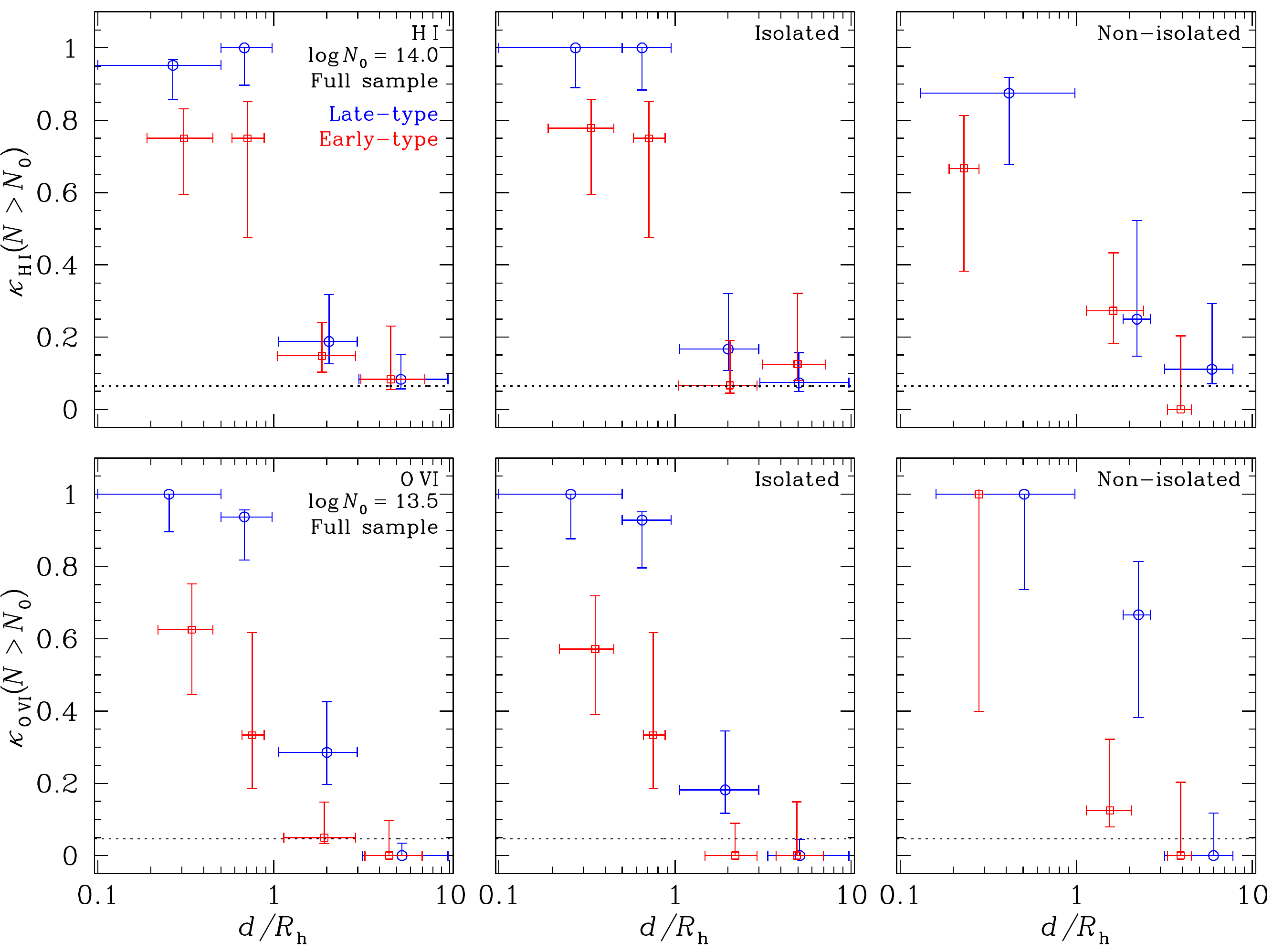}
	\caption{Covering fraction for strong H\,I and O\,VI absorption systems
	as a function of $d/R_{\rm h}$.
	The {\it top} panels show the covering fraction for 
	H\,I column density greater than $\log\,N_0({\rm H\,I})/{\rm cm}^{-2}=14.0$
	and the {\it bottom} panels show the covering fraction for
	O\,VI column density greater than
	$\log\,N_0({\rm O\,VI})/{\rm cm}^{-2}=13.5$.
	The late-type galaxy samples are shown in blue and early-type in red.
	The full eCGM and COS-Halos
	samples are shown on the {\it left}, isolated galaxies in the {\it middle}, and non-isolated
	ones on the {\it right}. The horizontal error bars mark the full range of galaxies within each
	bin and vertical error bars represent $68\%$ confidence intervals calculated using standard
	binomial statistics. We do not include non-detections with
	column density limits greater than
	$N_0$ in the covering fraction calculation.
	For both H\,I and O\,VI, we show the expected covering fraction for random sightlines
	based $\frac{d^2\mathcal{N}}{dNdz}$ measurements from \protect \cite{Danforth:2014}.}
	\label{figure:coveringfraction}

\end{figure*}

\section{The extent of H\,I and O\,VI absorption around galaxies}
\label{section:extent}
With the galaxy and absorption data described in
Sections \ref{section:galaxies} and \ref{section:absorption},
we characterize the H\,I and O\,VI column densities
of the eCGM and COS-Halos galaxy sightlines as
a function of projected distance
in Figure \ref{figure:d_Rh}.
The covering fractions for strong H\,I and O\,VI absorption
systems of $\log N({\rm H\,I})/{\rm cm}^{-2} > 14.0$
and $\log N({\rm O\,VI})/{\rm cm}^{-2} > 13.5$
are shown in Figure \ref{figure:coveringfraction}. 
The uncertainties in covering fractions are calculated using standard binomial statistics,
and we do not include galaxies with non-detections with upper limits greater than
$\log N({\rm H\,I})/{\rm cm}^{-2} > 13.5$
or $\log N({\rm O\,VI})/{\rm cm}^{-2} > 14.0$ in the calculations.
When multiple galaxies are associated with the same absorption system,
we show only the galaxy at the smallest $d/R_{\rm h}$.

\subsection{H\,I}
\label{section:HI}
Consistent with previous surveys \citep[e.g.][]{Chen:2001a},
we find a clear trend of decreasing H\,I
column density with increasing projected distance
(see the top panels of Figures \ref{figure:d_Rh} and \ref{figure:coveringfraction}).
Lyman-limit systems (LLS) and partial Lyman-limit systems
($\log\,N({\rm H\,I})/{\rm cm}^{-2} \gtrsim 16$) are common at $d < R_{\rm h}$,
but none are found at larger distances.
Considering the eCGM and COS-Halos samples together,
we identify strong H\,I systems with $\log\,N({\rm H\,I})/{\rm cm}^{-2} > 14$
around $48$ of $54$ galaxies at $d<R_{\rm h}$
indicating a mean covering fraction of
$\langle\kappa_{\rm H\,I}\rangle=0.89^{+0.03}_{-0.06}$.
At larger distances of $d/R_{\rm h}=1-3$ such H\,I absorption
systems are found for only $7$ of $43$ galaxies
($\langle\kappa_{\rm H\,I}\rangle=0.16^{+0.07}_{-0.04}$).
The H\,I covering fraction decreases to levels consistent with
expectations from coincidental absorption systems 
unrelated to the galaxy survey members at $d>3\,R_{\rm h}$.

To determine whether galaxy environment has an influence on
H\,I absorption far from galaxies, we show
the column densities and covering fractions for isolated
and non-isolated galaxies in the bottom middle and bottom
right panels of Figures \ref{figure:d_Rh} and \ref{figure:coveringfraction}.
Though the only two galaxies detected at $d>R_{\rm h}$
with $\log\,N({\rm H\,I})/{\rm cm}^{-2} \gtrsim 15$
have nearby neighbors, there is no clear enhancement in
H\,I absorption in the non-isolated sample.
A series of logrank tests over the range of $d/R_{\rm h}=1-3$ finds no
evidence for differential H\,I absorption between the isolated and non-isolated samples.
We note, however, that contamination from coincidental 
\lya\, absorbers in the IGM is non-neglegible.
This contamination may obscure a difference
in outer-halo H\,I absorption between the two samples.

Cross-correlation studies such as \cite{Chen:2009}
and \cite{Tejos:2014} indicate that strong H\,I
absorption systems primarily reside in the
gaseous halos of star-forming galaxies.
To investigate the possibility of enhanced H\,I
absorption in galaxies with recent star-formation
activity, we show late- and early-type galaxies
in blue and red respectively in Figures
\ref{figure:d_Rh} and \ref{figure:coveringfraction}.
Late-type galaxies exhibit excess H\,I absorption
relative to early-type galaxies
at projected distances of $d < R_{\rm h}$.
In particular, we find that $35$ of $36$ late-type
galaxies at $d < R_{\rm h}$ are associated
with H\,I systems of $\log\,N({\rm H\,I})/{\rm cm}^{-2} > 14.0$
indicating a mean covering fraction of
$\langle\kappa_{\rm H\,I}\rangle=0.97^{+0.01}_{-0.06}$
compared to $13$ of $18$ early-type galaxies 
($\langle\kappa_{\rm H\,I}\rangle=0.72^{+0.08}_{-0.12}$).
However, a comparison of the distributions of H\,I column densities
found at $d < R_{\rm h}$ around late- and early-type
galaxies is complicated by the poorly determined H\,I
column densities for the LLS and partial-LLS that are common
at small projected distances \cite[see discussion in][]{Tumlinson:2013}.

\begin{figure}
	\centering
	\includegraphics[scale=0.42]{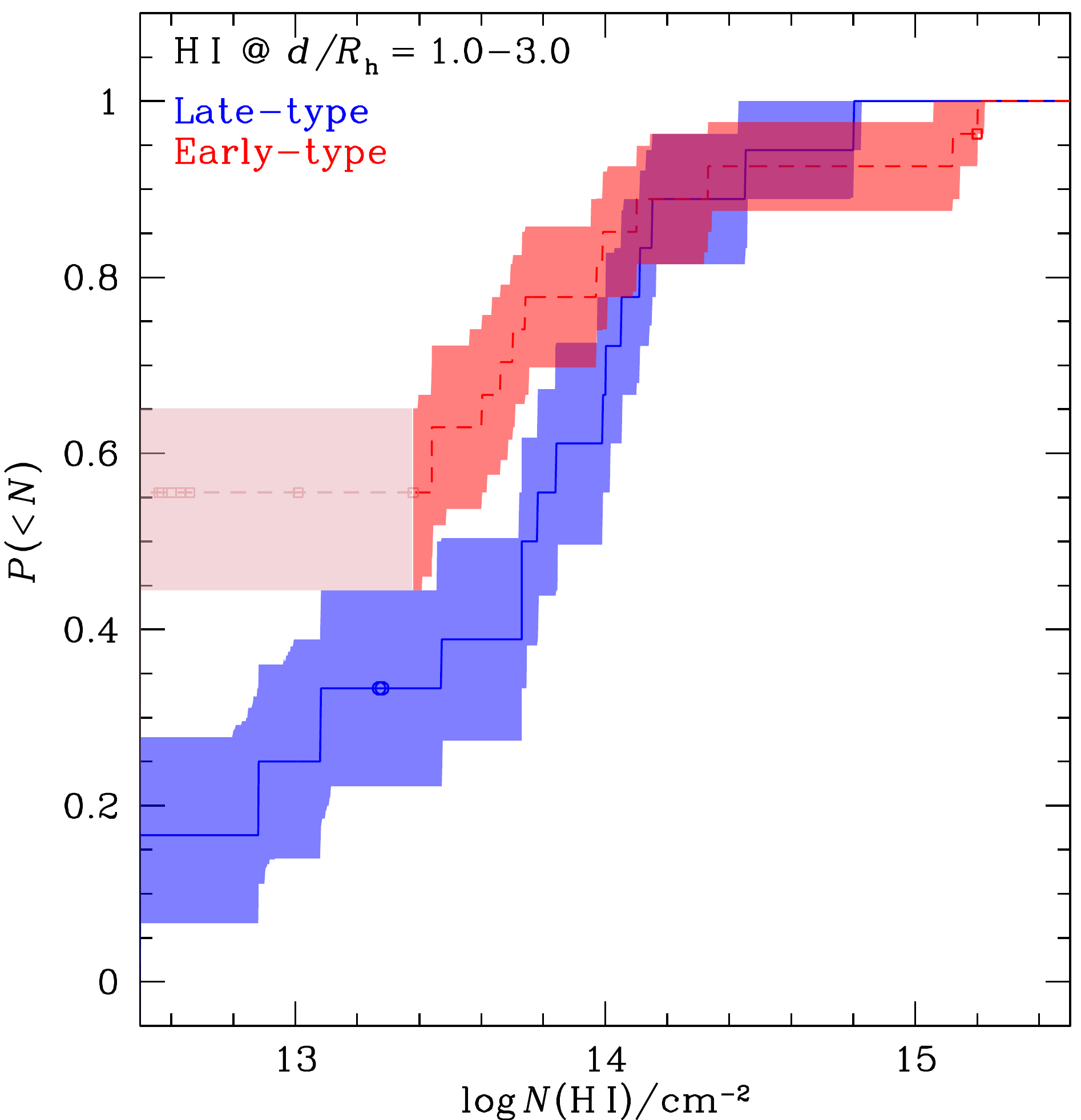}
	\caption{Cumulative fraction, $P$, of galaxies at $d/R_{\rm h}=1-3$
	with total H\,I column densities
	 no greater than $\log\,N({\rm H\,I})/{\rm cm}^{-2}$.
	The late-type galaxies are
	shown in blue solid line and early-type as red dashed line.
	The cumulative distributions are estimated using the Kaplan-Meier curve,
	an unbiased, non-parametric estimator of the cumulative distribution
	in the presence of upper-limits.
	The Kaplan-Meier estimator changes value at detections but not at non-detections.
	For clarity, non-detections with upper limits are shown as open symbols.
	The shaded bands represent $68\%$ confidence intervals
	including uncertainty due to both sample variance and column
	density measurement errors calculated with combined boot-strap
	and Monte-Carlo resampling.
	At column densities smaller than the highest upper limit that is less than
	the lowest detection, the Kaplan-Meier estimator represents an
	upper limit on the cumulative distribution.
	These column density ranges are shown in faded color.}
	\label{figure:kaplan}
\end{figure}

At larger distances of $d/R_{\rm h} = 1-3$
the H\,I covering fraction for systems with
$\log\,N({\rm H\,I})/{\rm cm}^{-2} > 14$
are similar for late- and early-type
galaxies. However, covering fractions are somewhat
insensitive to the underlying H\,I column density distribution.
To compare the H\,I column density distribution of
late- and early-type galaxies in the outer halo,
we perform a logrank test and find that the probability that
the column densities around the two galaxy classes at $d/R_{\rm h} = 1-3$
are drawn from the same underlying distribution to be $P=2\%$.
To investigate the nature of this likely differential H\,I absorption
around late- and early-type galaxies at large projected distances,
we compare the H\,I column densities of the two
samples using the Kaplan-Meier estimator,
an unbiased, non-parametric estimator of
cumulative distributions in the presence of
upper limits \citep[see ][]{Feigelson:1985}
in Figure \ref{figure:kaplan}.
We find excess H\,I absorption around
late-type galaxies driven primarily
by moderate strength absorption systems
of $\log\,N({\rm H\,I})/{\rm cm}^{-2} \approx 13.5$ to $14.0$.
Based on the Kaplan-Meier curves, we infer a median
H\,I column density of $\langle\log\,N({\rm H\,I})/{\rm cm}^{-2}\rangle \approx 13.8$
for late-type galaxies at $d/R_{\rm h} = 1-3$.
For early-type galaxies, the fraction of H\,I detections
at these distances is too small for a robust measurement
of the median, but we are able
to place an upper limit on the median H\,I column density of 
$\langle\log\,N({\rm H\,I})/{\rm cm}^{-2}\rangle \lesssim 13.4$.

\subsection{O\,VI}
\label{section:OVI}

Like H\,I, we find a clear anti-correlation between O\,VI column density
and projected distance (see the bottom left panel of  Figure \ref{figure:d_Rh}).
Strong O\,VI systems with $\log N({\rm O\,VI})/{\rm cm}^{-2} \approx 14.5-15.0$
are common at $d \lesssim R_{\rm h}$ but are rare at larger projected distances.
O\,VI systems with $\log N({\rm O\,VI})/{\rm cm}^{-2} \approx 13.5-14.0$
are detected out to $d/R_{\rm h}=2.3$, but
at larger distances, no such strong O\,VI systems are found.
Considering the eCGM and COS-Halos samples as a whole,
we find O\,VI absorption with $\log N({\rm O\,VI})/{\rm cm}^{-2} > 13.5$
for $36$ of $42$ galaxies at $d<R_{\rm h}$
indicating a mean covering fraction of
$\langle\kappa_{\rm O\,VI}\rangle=0.86^{+0.04}_{-0.07}$.
Isolated galaxies exhibit O\,VI absorption out to $d \approx R_{\rm h}$,
beyond the ``metal-boundary'' seen in C\,IV and lower ions
at $d\approx 0.7\,R_{\rm h}$.
At $d/ R_{\rm h} = 1-3$ strong O\,VI absorption
systems are found for only $5$ of $34$ galaxies
($\langle\kappa_{\rm O\,VI}\rangle=0.15^{+0.08}_{-0.04}$).
At larger distances of $d > 3\,R_{\rm h}$, the O\,VI covering
fraction decreases to levels consistent with expectations
from coincidental absorption systems unrelated to the galaxy survey members
(see the bottom left panel of Figure \ref{figure:coveringfraction}).

The high completeness of the galaxy survey enables a detailed investigation
of possible environmental dependence of extended gas around galaxies, and
we show O\,VI column density and covering fraction versus $d/R_{\rm h}$
for isolated and non-isolated galaxies separately in the bottom middle and
bottom right panels of Figures \ref{figure:d_Rh} and \ref{figure:coveringfraction}.
The decrease in incidence of strong O\,VI systems at $d\approx R_{\rm h}$
is particularly sharp among isolated galaxies.
Of the $51$ isolated galaxies at $d>R_{\rm h}$ in the sample
we detect $\log N({\rm O\,VI})/{\rm cm}^{-2} > 13.5$ absorption
systems for only two galaxies (at $d/R_{\rm h}=1.1$ and $1.2$)
indicating a mean covering fraction of 
$\kappa_{\rm O\,VI}=0.04^{+0.05}_{-0.01}$.
In contrast, three non-isolated galaxies at $d/R_{\rm h}=1.3, 1.8, $ and $2.3$  
are found to have absorption systems with $\log N({\rm O\,VI})/{\rm cm}^{-2} > 13.5$
out of a sample of $24$ such galaxies at $d>R_{\rm h}$
($\kappa_{\rm O\,VI}=0.13^{+0.10}_{-0.04}$).
This suggests that the O\,VI absorption around non-isolated
galaxies is more extended than around isolated galaxies.
 
To evaluate the possibility that non-isolated galaxies have
more extended O\,VI absorbing gas, we compare
the O\,VI column density distributions of the isolated
and non-isolated galaxy samples at $d>1.5\,R_{\rm h}$.
A logrank test finds that the probability that the O\,VI column densities
of the isolated and non-isolated galaxies at $d>1.5\,R_{\rm h}$ 
are drawn from the same underlying distribution to be $P<1\%$.
The results of the logrank test are insensitive to the choice of 
$d>1.5\,R_{\rm h}$ cut-off and return a result of $P<1\%$
over the full range of cut-off distances between $1.2$ and $1.8\,R_{\rm h}$
and $P<2\%$ between $1.1$ and $2.2\,R_{\rm h}$.
These findings indicate that galaxies with nearby neighbors
exhibit more extended O\,VI absorbing gas
than isolated galaxies.

Though the completeness of our galaxy survey is high, it is
possible that the detections of O\,VI at large distances from
non-isolated galaxies could be due to the presence of an
additional neighbor closer to the sightline that was not observed
in our survey. However, the survey incompleteness applies to the isolated
and non-isolated galaxies equally. In addition, the galaxy survey completeness
levels increase at small projected distances from the quasar sightline
\citep[see Figure 2 of][]{Johnson:2013}.
In particular, at the redshifts
and distances of the non-isolated galaxies detected in O\,VI absorption
at $d>R_{\rm h}$, our galaxy survey data are $>90\%$ complete for
galaxies as faint as $0.1\,L_*$.

Previous surveys \citep[e.g.][]{Chen:2009, Tumlinson:2011}
found that late-type galaxies exhibit enhanced O\,VI absorption
relative to more massive early-type galaxies at $d\lesssim 150$ kpc.
We confirm that late-type galaxies show enhanced O\,VI absorption
at $d<R_{\rm h}$ relative to early-type ones.
However, O\,VI absorption systems with 
$\log N({\rm O\,VI})/{\rm cm}^{-2} > 13.5$
are found for six of eleven early-type
galaxies at $d<R_{\rm h}$
indicating a non-negligible covering fraction 
of $\langle\kappa_{\rm O\,VI}\rangle \approx 0.55^{+0.13}_{-0.14}$.
At larger distances of $d/R_{\rm h}=1-3$
we find O\,VI absorption with $\log N({\rm O\,VI})/{\rm cm}^{-2} > 13.5$
for four of $14$ late-type galaxies 
$\langle\kappa_{\rm O\,VI}\rangle \approx 0.29^{+0.14}_{-0.09}$
compared to  one of $19$ early-type galaxies
$\langle\kappa_{\rm O\,VI}\rangle \approx 0.05^{+0.10}_{-0.02}$.
Interestingly, the only early-type galaxy exhibiting a
strong O\,VI system at $d>R_{\rm h}$ is the
brightest member of a maxBCG cluster \citep{Koester:2007}
at $d=1.3\,R_{\rm h}$ from the quasar sightline.
A logrank test finds that the probability that the O\,VI column densities
found at $d/R_{\rm h}=1-3$ from late- and early-type
galaxies are drawn from the same underlying distribution to be $P=1\%$.
We note that the O\,VI excess in the outer halos
of late-type galaxies is driven by absorption systems
with $\log N({\rm O\,VI})/{\rm cm}^{-2} \approx 13.5-14.0$
while the excess at smaller projected distances is driven by
stronger systems with $\log N({\rm O\,VI})/{\rm cm}^{-2} \approx 14.5$.

\section{Discussion}
\label{section:discussion}
Using a highly complete survey ($\gtrsim80\%$) of faint galaxies ($L>0.1\,L_*$)
at $z<0.4$ and $d<500$ kpc in the fields of four
COS quasar sightlines with high signal-to-noise
spectra available in the {\it HST} archive, we have searched
for the key features that determine the extended gaseous
properties of galaxy halos in H\,I and O\,VI absorption.
Our main findings are:
\begin{enumerate}
\item galaxies with nearby neighbors exhibit enhanced O\,VI absorption at $d>R_{\rm h}$ relative to isolated galaxies;
\item among isolated galaxies, O\,VI absorption extends to $d \approx R_{\rm h}$, beyond the extent observed in lower ionization species such as Si\,III and C\,IV; and
\item late-type galaxies exhibit enhanced H\,I and O\,VI absorption beyond characteristic halo radii at $d/R_{\rm h}=1-3$.
\end{enumerate}

\subsection{Possible environmental effect in distributing heavy elements to $d>R_{\rm h}$}
Among isolated galaxies without neighbors within $500$ kpc and with stellar mass ratio greater than one-third,
we find that O\,VI absorption is confined to within $d<1.2\,R_{\rm h}$.
None of the $18$ isolated galaxies at $d/R_{\rm h}=1.2-3.0$ in our survey
are found to have O\,VI absorption in the COS quasar spectrum.
In contrast, three out of ten galaxies with nearby neighbors
probed at $d/R_{\rm h}=1.2-3.0$ are found to exhibit O\,VI absorption
with $\log\,N({\rm O\,VI})/{\rm cm}^{-2} > 13.5$.
On the other hand, no evidence for differential H\,I absorption is found between
the isolated and non-isolated galaxy samples.
These findings suggest that galaxy interactions play a key role in distributing heavy
elements to large distances from galaxy centers, well beyond the enriched
gaseous halos that surround individual galaxies.

Galaxy interactions can produce heavy elements
at large projected distances through tidal stripping
during satellite accretion \citep[e.g.][]{Wang:1993}
or ram-pressure stripping \citep[e.g.][]{Gunn:1972}.
Recently, {\it HST} images of galaxy clusters have
revealed that ram pressure stripping can remove
and compress copious amounts of gas from in-falling galaxies
as evidenced by vigorous starbursts in debris trails \citep[][]{Ebeling:2014}.
Moreover, a galaxy survey in the field of an ultra-strong Mg\,II absorption system
near a luminous red galaxy revealed a group of evolved galaxies
with no evidence of recent star-formation activity \citep{Gauthier:2013}.
The lack of star-formation in the group and kinematics of the Mg\,II absorption
are most readily explained if the Mg\,II absorber arises from stripped gas.
Finally, in \cite{Johnson:2014} we presented the discovery of a
``transparent'' sightline with no strong H\,I, Mg\,II or O\,VI absorption
at $d<20$ kpc from a pair of strongly interacting galaxies separated by
a projected distance of $9$ kpc. The lack of strong absorption systems
detected at such small projected distance from this galaxy pair
can be explained if the inner halo gas of the two galaxies
has been stripped to larger distances during the galaxy interaction.

\subsection{Comparison with Prochaska et al. 2011}
\begin{figure}
	\centering
	\includegraphics[scale=0.42]{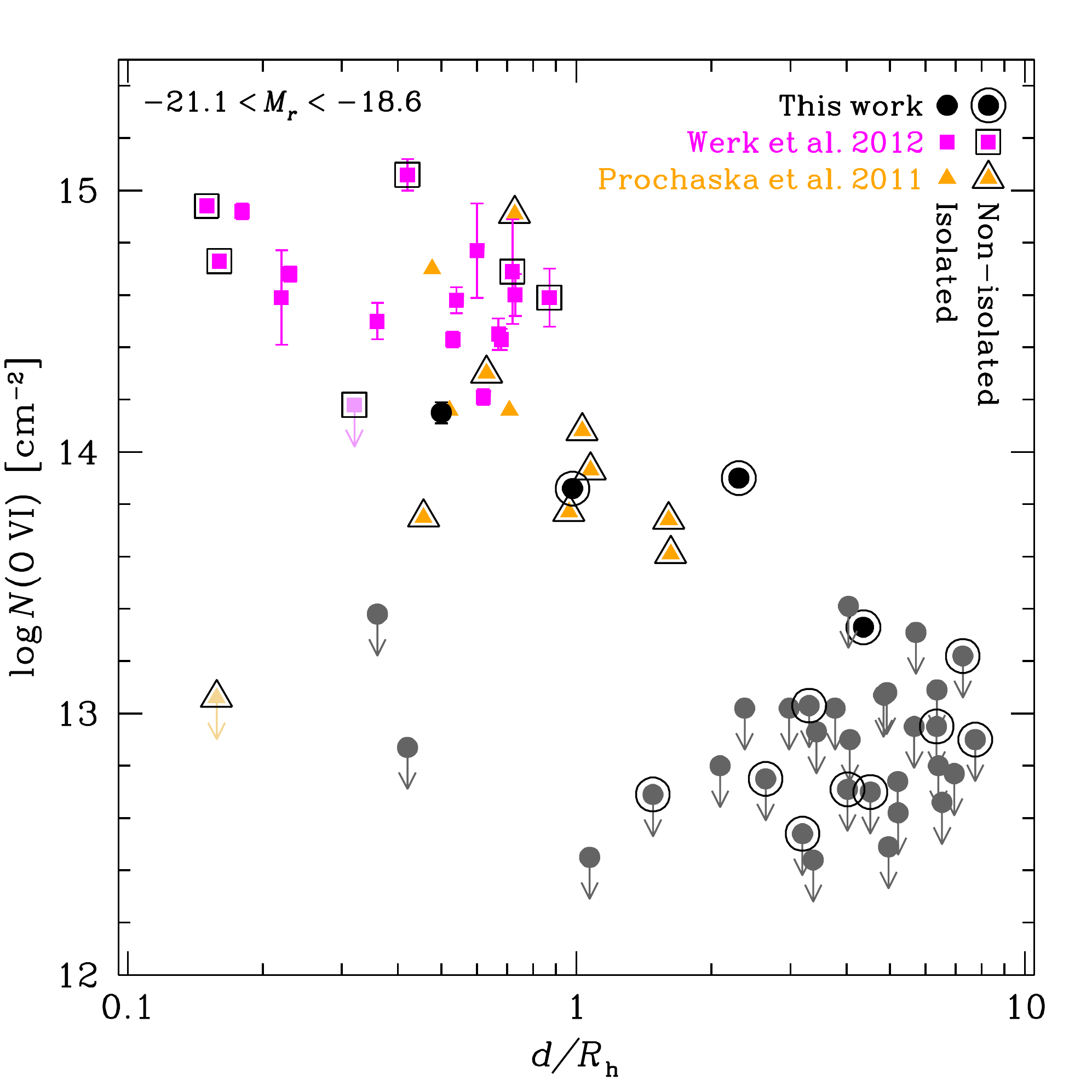}
	\caption{O\,VI column density versus $d/R_{\rm h}$ for sub-$L_*$ galaxies of $-21.1<M_r<-18.6$. Galaxies from the eCGM survey are shown as black circles, those from \protect \cite{Werk:2012} as magenta squares, and those from  \protect \cite{Prochaska:2011} as orange triangles. Non-isolated galaxies are outlined by large black symbols.}
	\label{figure:Prochaska2011}
\end{figure}

\cite{Prochaska:2011} found that galaxies at
$z\lesssim0.1$ and of $-21.1 < M_r < -18.6$ exhibit
near unity covering fraction at $d<300$ kpc and concluded
that such ``sub-$L_*$'' galaxies possess O\,VI bearing gaseous
halos that extend to $d\approx3\,R_{\rm h}$.
Our highly complete galaxy survey data shed new light on these
observations, indicating that O\,VI absorption at such large
distances from galaxy halos arise in systems of multiple galaxies.
Indeed, a literature search of the NASA Extragalactic database
for spectroscopic neighbors reveals that all of the galaxies
of $-21.1 < M_r < -18.6$ from \cite{Prochaska:2011}
with O\,VI absorption at $d\gtrsim R_{\rm h}$
have neighbors within $d=500$ kpc and satisfy our definition of a
non-isolated galaxy (see Figure \ref{figure:Prochaska2011}).
The data from \cite{Prochaska:2011} are therefore consistent
with our conclusion that isolated galaxies exhibit low covering fractions
at $d \gtrsim 1.2\,R_{\rm h}$ and support the possibility that 
galaxy interactions are effective at producing O\,VI absorbing gas at large
projected distances.

\subsection{O\,VI absorption in isolated galaxies beyond the ``metal-boundary'' at $d/R_{\rm h}=0.7$}
Even among isolated galaxies, O\,VI absorption extends
beyond the ``metal-boundary'' observed
by \cite{Liang:2014} at $d=0.7\,R_{\rm h}$ 
in lower-ionization state,
enriched gas traced by absorbers such as Si\,II, Si\,III, C\,II, and C\,IV.
Specifically, we find O\,VI absorption
for seven of ten isolated galaxies
at $d/R_{\rm h}=0.7$ to $1.2$.
The presence of O\,VI
absorption in the outer regions of isolated galaxy halos
indicates some heavy element enrichment at large radii
and that highly ionized, O\,VI bearing clouds can form or survive
in the outer halo while Si\,II, Si\,III, and C\,IV bearing clouds cannot.
Using the mean absorption found in a stack
of isolated galaxies in the eCGM and COS-Halos samples at
$d/R_{\rm h}=0.7-1.2$, we estimate the mean metallicity
of these clouds.
In the stack, we measure absorption of
$\log N({\rm H\,I})/{\rm cm}^{-2} = 14.4$,
$\log N({\rm O\,VI})/{\rm cm}^{-2} = 13.9$,
and place an upper limit on C\,III
absorption of $\log N({\rm C\,III})/{\rm cm}^{-2} < 12.6$.
Assuming photoionization equilibrium models
calculated with Cloudy \citep{Ferland:1998} Version 10.0,
a photoionization background from \cite{Haardt:2012}
at $z=0.2$, and solar abundance patterns,
these mean column densities are consistent with
gas in photoionization equilibrium with ionization
parameter $\log\,U = -1.1$ and
a metallicity of one-tenth solar.
The O\,VI and H\,I bearing gas
may not be co-spatial allowing the ionization
parameter of the H\,I gas to be lower.
If this is the case, then
the upper limit on C\,III absorption places
stricter upper limits on the mean gas metallicity.
We therefore place an upper limit of
one-tenth solar on the mean metallicity
of outer-halo H\,I gas clouds at $d\approx\,R_{\rm h}$.
This is consistent with the metallicity upper limit of
one-tenth solar found by \cite{Liang:2014}
based on C\,IV non-detections at $d>0.8\,R_{\rm h}$.

\subsection{Differential H\,I and O\,VI absorption between late- versus early-type galaxies}
We find excess H\,I and O\,VI absorption around late-type
galaxies relative to early-type ones
at $d/R_{\rm h}=1-3$ with $99\%$ and $98\%$ confidence respectively,
extending previous results from \cite{Chen:2009}, \cite{Tumlinson:2011}, and \cite{Tejos:2014} to larger distances.
The O\,VI excess around late-type galaxies
could be explained if the ionization mechanisms that produce
O\,VI absorbing gas are the result of star-formation
(e.g. photoionization from a young stellar population).
However, this is not consistent with the observed H\,I
excess unless the ionization mechanism is effective
at producing highly ionized heavy element species but
ineffective at ionizing hydrogen (e.g. Auger ionization).

Higher levels of H\,I  and O\,VI absorption around
late-type galaxies could be simultaneously explained if the absorption
systems trace starburst driven outflows.
However, simulations from \cite{Oppenheimer:2008}
that incorporate strong, momentum-driven winds
find that these outflows are characterized by a
turnaround radius of $R_{\rm turn} = 80 \pm 20$ kpc
with only weak dependence on mass and redshift at $z<1$.
For the late-type galaxies in our survey, this corresponds to only
$R_{\rm turn}/R_{\rm h} \approx 0.4$, far less than $d/R_{\rm h}=1-3$.
Alternatively, the excess H\,I and O\,VI absorption at large distances from late-type
galaxies can be explained if H\,I, O\,VI, and star-formation
are observational signatures of halos in cosmic
environments with multi-phase gaseous reservoirs.

The late-type members of the eCGM and COS-Halos
galaxy samples are systematically less massive than
the early-type members with mean stellar masses of 
$\log\,\langle M_{*}/M_\odot \rangle=10.2, 10.9$
corresponding to inferred halo masses of
$\log\,\langle M_{h}/M_\odot \rangle =11.7, 12.6$ respectively.
The mean doppler parameter of the H\,I absorption
components detected at $d/R_{\rm h}=1-3$
from eCGM galaxies is $b=40$ \kms\, which indicates
that the H\,I absorbers trace cool-warm, $T\lesssim10^5$ K gas.
Our survey data are therefore consistent with
a decreased incidence of cool-warm gas around more massive halos.
This extends previous results from \cite{Gauthier:2010}
and \cite{Yoon:2012} who found reduced cool-warm gas absorption
in $>10^{13} M_\odot$ mass halos relative to less massive ones.
A larger sample of early-type galaxies probed in absorption
is required to differentiate the effects of recent star-formation
activity and mass.

\section*{Acknowledgements}
We thank
Michael Rauch, Gwen Rudie, Juna Kollmeier,
Don York, Jean-Ren\'{e} Gauthier, Rik Williams,
Cameron Jia Liang, Daniel D. Kelson, and Andrey Kravtsov
for many helpful discussions.
SDJ gratefully acknowledges support from
The Brinson Foundation and
The Observatories of the Carnegie Institute of Washington
who generously supported and hosted his studies over the last year.

This paper includes data gathered with the 6.5 meter Magellan Telescopes located at Las Campanas Observatory, Chile.

Based on observations made with the NASA/ESA Hubble Space Telescope, obtained from the Data Archive at the Space Telescope Science Institute, which is operated by the Association of Universities for Research in Astronomy, Inc., under NASA contract NAS 5-26555.

Funding for SDSS-III has been provided by the Alfred P. Sloan Foundation, the Participating Institutions, the National Science Foundation, and the U.S. Department of Energy Office of Science. The SDSS-III web site is http://www.sdss3.org/.
SDSS-III is managed by the Astrophysical Research Consortium for the Participating Institutions of the SDSS-III Collaboration including the University of Arizona, the Brazilian Participation Group, Brookhaven National Laboratory, Carnegie Mellon University, University of Florida, the French Participation Group, the German Participation Group, Harvard University, the Instituto de Astrofisica de Canarias, the Michigan State/Notre Dame/JINA Participation Group, Johns Hopkins University, Lawrence Berkeley National Laboratory, Max Planck Institute for Astrophysics, Max Planck Institute for Extraterrestrial Physics, New Mexico State University, New York University, Ohio State University, Pennsylvania State University, University of Portsmouth, Princeton University, the Spanish Participation Group, University of Tokyo, University of Utah, Vanderbilt University, University of Virginia, University of Washington, and Yale University.

This research made use of NASA's Astrophysics Data System (ADS)
and the NASA/IPAC Extragalactic Database (NED) which is operated by the Jet Propulsion Laboratory, California Institute of Technology, under contract with the National Aeronautics and Space Administration.

Funding for PRIMUS is provided by NSF (AST-0607701, AST-0908246, AST-0908442, AST-0908354) and NASA (Spitzer-1356708, 08-ADP08-0019, NNX09AC95G).

This research draws upon data provided by Brian Keeney as distributed by the NOAO Science Archive. NOAO is operated by the Association of Universities for Research in Astronomy (AURA) under cooperative agreement with the National Science Foundation.

\footnotesize{
\bibliographystyle{mn} 
\bibliography{biblio}

\begin{thebibliography}{65}
\expandafter\ifx\csname natexlab\endcsname\relax\def\natexlab#1{#1}\fi

\bibitem[{{Agertz} \& {Kravtsov}(2014)}]{Agertz:2014}
{Agertz} O., {Kravtsov} A.~V., 2014, ArXiv e-prints, 1404.2613

\bibitem[{{Ahn} {et~al.}(2014){Ahn}, {Alexandroff}, {Allende Prieto}, {Anders},
  {Anderson}, \& {\etal}}]{Ahn:2014}
{Ahn} C.~P., {Alexandroff} R., {Allende Prieto} C., {Anders} F., {Anderson}
  S.~F., {\etal}, 2014, ApJS, 211, 17

\bibitem[{{Behroozi} {et~al.}(2013){Behroozi}, {Wechsler}, \&
  {Conroy}}]{Behroozi:2013}
{Behroozi} P.~S., {Wechsler} R.~H., {Conroy} C., 2013, ApJ, 770, 57

\bibitem[{{Bernardi} {et~al.}(2013){Bernardi}, {Meert}, {Sheth}, {Vikram},
  {Huertas-Company}, {Mei}, \& {Shankar}}]{Bernardi:2013}
{Bernardi} M., {Meert} A., {Sheth} R.~K., {Vikram} V., {Huertas-Company} M.,
  {Mei} S., {Shankar} F., 2013, MNRAS, 436, 697

\bibitem[{{Bertin} \& {Arnouts}(1996)}]{Bertin:1996}
{Bertin} E., {Arnouts} S., 1996, Astronomy and Astrophysics, Supplement, 117,
  393

\bibitem[{{Blanton} \& {Roweis}(2007)}]{Blanton:2007}
{Blanton} M.~R., {Roweis} S., 2007, AJ, 133, 734

\bibitem[{{Bordoloi} {et~al.}(2011){Bordoloi}, {Lilly}, {Knobel}, {Bolzonella},
  {Kampczyk}, , \& {\etal}}]{Bordoloi:2011}
{Bordoloi} R., {Lilly} S.~J., {Knobel} C., {Bolzonella} M., {Kampczyk} P., ,
  {\etal}, 2011, ApJ, 743, 10

\bibitem[{{Bordoloi} {et~al.}(2014){Bordoloi}, {Tumlinson}, {Werk},
  {Oppenheimer}, {Peeples}, {Prochaska}, {Tripp}, {Katz}, {Dav{\'e}}, {Fox},
  {Thom}, {Ford}, {Weinberg}, {Burchett}, \& {Kollmeier}}]{Bordoloi:2014}
{Bordoloi} R., {Tumlinson} J., {Werk} J.~K., {Oppenheimer} B.~D., {Peeples}
  M.~S., {Prochaska} J.~X., {Tripp} T.~M., {Katz} N., {Dav{\'e}} R., {Fox}
  A.~J., {Thom} C., {Ford} A.~B., {Weinberg} D.~H., {Burchett} J.~N.,
  {Kollmeier} J.~A., 2014, ApJ, 796, 136

\bibitem[{{Borthakur} {et~al.}(2013){Borthakur}, {Heckman}, {Strickland},
  {Wild}, \& {Schiminovich}}]{Borthakur:2013}
{Borthakur} S., {Heckman} T., {Strickland} D., {Wild} V., {Schiminovich} D.,
  2013, ApJ, 768, 18

\bibitem[{{Bowen} {et~al.}(1995){Bowen}, {Blades}, \& {Pettini}}]{Bowen:1995}
{Bowen} D.~V., {Blades} J.~C., {Pettini} M., 1995, ApJ, 448, 634

\bibitem[{{Bryan} \& {Norman}(1998)}]{Bryan:1998}
{Bryan} G.~L., {Norman} M.~L., 1998, ApJ, 495, 80

\bibitem[{{Carswell} {et~al.}(1987){Carswell}, {Webb}, {Baldwin}, \&
  {Atwood}}]{Carswell:1987}
{Carswell} R.~F., {Webb} J.~K., {Baldwin} J.~A., {Atwood} B., 1987, ApJ, 319,
  709

\bibitem[{{Cen}(2013)}]{Cen:2013}
{Cen} R., 2013, ApJ, 770, 139

\bibitem[{{Chabrier}(2003)}]{Chabrier:2003}
{Chabrier} G., 2003, PASP, 115, 763

\bibitem[{{Chen} {et~al.}(2010){Chen}, {Helsby}, {Gauthier}, {Shectman},
  {Thompson}, \& {Tinker}}]{Chen:2010a}
{Chen} H.-W., {Helsby} J.~E., {Gauthier} J.-R., {Shectman} S.~A., {Thompson}
  I.~B., {Tinker} J.~L., 2010, ApJ, 714, 1521

\bibitem[{{Chen} {et~al.}(1998){Chen}, {Lanzetta}, {Webb}, \&
  {Barcons}}]{Chen:1998}
{Chen} H.-W., {Lanzetta} K.~M., {Webb} J.~K., {Barcons} X., 1998, ApJ, 498, 77

\bibitem[{{Chen} {et~al.}({2001}){Chen}, {Lanzetta}, {Webb}, \&
  {Barcons}}]{Chen:2001a}
---, {2001}, ApJ, 559, 654

\bibitem[{{Chen} \& {Mulchaey}(2009)}]{Chen:2009}
{Chen} H.-W., {Mulchaey} J.~S., 2009, ApJ, 701, 1219

\bibitem[{{Coil} {et~al.}(2011){Coil}, {Blanton}, {Burles}, {Cool},
  {Eisenstein}, \& {\etal}}]{Coil:2011}
{Coil} A.~L., {Blanton} M.~R., {Burles} S.~M., {Cool} R.~J., {Eisenstein}
  D.~J., {\etal}, 2011, ApJ, 741, 8

\bibitem[{{Danforth} {et~al.}(2014){Danforth}, {Tilton}, {Shull}, {Keeney},
  {Stevans}, \& {\etal}}]{Danforth:2014}
{Danforth} C.~W., {Tilton} E.~M., {Shull} J.~M., {Keeney} B.~A., {Stevans} M.,
  {\etal}, 2014, ArXiv e-prints, 1402.2655

\bibitem[{{Dressler} {et~al.}(2011){Dressler}, {Bigelow}, {Hare}, {Sutin},
  {Thompson}, \& {\etal}}]{Dressler:2011}
{Dressler} A., {Bigelow} B., {Hare} T., {Sutin} B., {Thompson} I., {\etal},
  2011, PASP, 123, 288

\bibitem[{{Ebeling} {et~al.}(2014){Ebeling}, {Stephenson}, \&
  {Edge}}]{Ebeling:2014}
{Ebeling} H., {Stephenson} L.~N., {Edge} A.~C., 2014, ApJL, 781, L40

\bibitem[{{Feigelson} \& {Nelson}(1985)}]{Feigelson:1985}
{Feigelson} E.~D., {Nelson} P.~I., 1985, ApJ, 293, 192

\bibitem[{{Ferland} {et~al.}(1998){Ferland}, {Korista}, {Verner}, {Ferguson},
  {Kingdon}, \& {Verner}}]{Ferland:1998}
{Ferland} G.~J., {Korista} K.~T., {Verner} D.~A., {Ferguson} J.~W., {Kingdon}
  J.~B., {Verner} E.~M., 1998, PASP, 110, 761

\bibitem[{{Ford} {et~al.}(2013){Ford}, {Oppenheimer}, {Dav{\'e}}, {Katz},
  {Kollmeier}, \& {Weinberg}}]{Ford:2013}
{Ford} A.~B., {Oppenheimer} B.~D., {Dav{\'e}} R., {Katz} N., {Kollmeier} J.~A.,
  {Weinberg} D.~H., 2013, MNRAS, 432, 89

\bibitem[{{Gauthier}(2013)}]{Gauthier:2013}
{Gauthier} J.-R., 2013, MNRAS, 432, 1444

\bibitem[{{Gauthier} {et~al.}(2010){Gauthier}, {Chen}, \&
  {Tinker}}]{Gauthier:2010}
{Gauthier} J.-R., {Chen} H.-W., {Tinker} J.~L., 2010, ApJ, 716, 1263

\bibitem[{{Green} {et~al.}(2012){Green}, {Froning}, {Osterman}, {Ebbets},
  {Heap}, \& {\etal}}]{Green:2012}
{Green} J.~C., {Froning} C.~S., {Osterman} S., {Ebbets} D., {Heap} S.~H.,
  {\etal}, 2012, ApJ, 744, 60

\bibitem[{{Gunn} \& {Gott}(1972)}]{Gunn:1972}
{Gunn} J.~E., {Gott} III J.~R., 1972, ApJ, 176, 1

\bibitem[{{Haardt} \& {Madau}(2012)}]{Haardt:2012}
{Haardt} F., {Madau} P., 2012, ApJ, 746, 125

\bibitem[{{Hummels} {et~al.}(2013){Hummels}, {Bryan}, {Smith}, \&
  {Turk}}]{Hummels:2013}
{Hummels} C.~B., {Bryan} G.~L., {Smith} B.~D., {Turk} M.~J., 2013, MNRAS, 430,
  1548

\bibitem[{{Johnson} {et~al.}(2013){Johnson}, {Chen}, \&
  {Mulchaey}}]{Johnson:2013}
{Johnson} S.~D., {Chen} H.-W., {Mulchaey} J.~S., 2013, MNRAS, 434, 1765

\bibitem[{{Johnson} {et~al.}(2014){Johnson}, {Chen}, {Mulchaey}, {Tripp},
  {Prochaska}, \& {Werk}}]{Johnson:2014}
{Johnson} S.~D., {Chen} H.-W., {Mulchaey} J.~S., {Tripp} T.~M., {Prochaska}
  J.~X., {Werk} J.~K., 2014, MNRAS, 438, 3039

\bibitem[{{Koester} {et~al.}(2007){Koester}, {McKay}, {Annis}, {Wechsler},
  {Evrard}, {Bleem}, {Becker}, {Johnston}, {Sheldon}, {Nichol}, {Miller},
  {Scranton}, {Bahcall}, {Barentine}, {Brewington}, {Brinkmann}, {Harvanek},
  {Kleinman}, {Krzesinski}, {Long}, {Nitta}, {Schneider}, {Sneddin}, {Voges},
  \& {York}}]{Koester:2007}
{Koester} B.~P., {McKay} T.~A., {Annis} J., {Wechsler} R.~H., {Evrard} A.,
  {Bleem} L., {Becker} M., {Johnston} D., {Sheldon} E., {Nichol} R., {Miller}
  C., {Scranton} R., {Bahcall} N., {Barentine} J., {Brewington} H., {Brinkmann}
  J., {Harvanek} M., {Kleinman} S., {Krzesinski} J., {Long} D., {Nitta} A.,
  {Schneider} D.~P., {Sneddin} S., {Voges} W., {York} D., 2007, ApJ, 660, 239

\bibitem[{{Kravtsov} {et~al.}(2014){Kravtsov}, {Vikhlinin}, \&
  {Meshscheryakov}}]{Kravtsov:2014}
{Kravtsov} A., {Vikhlinin} A., {Meshscheryakov} A., 2014, ArXiv e-prints

\bibitem[{{Liang} \& {Chen}(2014)}]{Liang:2014}
{Liang} C.-J., {Chen} H.-W., 2014, ArXiv e-prints

\bibitem[{{Loveday} {et~al.}(2012){Loveday}, {Norberg}, {Baldry}, {Driver},
  {Hopkins}, {Peacock}, {Bamford}, {Liske}, {Bland-Hawthorn}, {Brough},
  {Brown}, {Cameron}, {Conselice}, {Croom}, {Frenk}, {Gunawardhana}, {Hill},
  {Jones}, {Kelvin}, {Kuijken}, {Nichol}, {Parkinson}, {Phillipps}, {Pimbblet},
  {Popescu}, {Prescott}, {Robotham}, {Sharp}, {Sutherland}, {Taylor}, {Thomas},
  {Tuffs}, {van Kampen}, \& {Wijesinghe}}]{Loveday:2012}
{Loveday} J., {Norberg} P., {Baldry} I.~K., {Driver} S.~P., {Hopkins} A.~M.,
  {Peacock} J.~A., {Bamford} S.~P., {Liske} J., {Bland-Hawthorn} J., {Brough}
  S., {Brown} M.~J.~I., {Cameron} E., {Conselice} C.~J., {Croom} S.~M., {Frenk}
  C.~S., {Gunawardhana} M., {Hill} D.~T., {Jones} D.~H., {Kelvin} L.~S.,
  {Kuijken} K., {Nichol} R.~C., {Parkinson} H.~R., {Phillipps} S., {Pimbblet}
  K.~A., {Popescu} C.~C., {Prescott} M., {Robotham} A.~S.~G., {Sharp} R.~G.,
  {Sutherland} W.~J., {Taylor} E.~N., {Thomas} D., {Tuffs} R.~J., {van Kampen}
  E., {Wijesinghe} D., 2012, MNRAS, 420, 1239

\bibitem[{{Maller} {et~al.}(2009){Maller}, {Berlind}, {Blanton}, \&
  {Hogg}}]{Maller:2009}
{Maller} A.~H., {Berlind} A.~A., {Blanton} M.~R., {Hogg} D.~W., 2009, ApJ, 691,
  394

\bibitem[{{Maller} \& {Bullock}(2004)}]{Maller:2004}
{Maller} A.~H., {Bullock} J.~S., 2004, MNRAS, 355, 694

\bibitem[{{Mathes} {et~al.}(2014){Mathes}, {Churchill}, {Kacprzak}, {Nielsen},
  {Trujillo-Gomez}, {Charlton}, \& {Muzahid}}]{Mathes:2014}
{Mathes} N.~L., {Churchill} C.~W., {Kacprzak} G.~G., {Nielsen} N.~M.,
  {Trujillo-Gomez} S., {Charlton} J., {Muzahid} S., 2014, ArXiv e-prints

\bibitem[{{McConnachie} {et~al.}(2009){McConnachie}, {Patton}, {Ellison}, \&
  {Simard}}]{McConnachie:2009}
{McConnachie} A.~W., {Patton} D.~R., {Ellison} S.~L., {Simard} L., 2009, MNRAS,
  395, 255

\bibitem[{{Montero-Dorta} \& {Prada}(2009)}]{Dorta:2009}
{Montero-Dorta} A.~D., {Prada} F., 2009, MNRAS, 399, 1106

\bibitem[{{Moster} {et~al.}(2010){Moster}, {Somerville}, {Maulbetsch}, {van den
  Bosch}, {Macci{\`o}}, {Naab}, \& {Oser}}]{Moster:2010}
{Moster} B.~P., {Somerville} R.~S., {Maulbetsch} C., {van den Bosch} F.~C.,
  {Macci{\`o}} A.~V., {Naab} T., {Oser} L., 2010, ApJ, 710, 903

\bibitem[{{Murray} {et~al.}(2011){Murray}, {M{\'e}nard}, \&
  {Thompson}}]{Murray:2011}
{Murray} N., {M{\'e}nard} B., {Thompson} T.~A., 2011, ApJ, 735, 66

\bibitem[{{Oppenheimer} \& {Dav{\'e}}(2008)}]{Oppenheimer:2008}
{Oppenheimer} B.~D., {Dav{\'e}} R., 2008, MNRAS, 387, 577

\bibitem[{{Prochaska} {et~al.}(2011){Prochaska}, {Weiner}, {Chen}, {Mulchaey},
  \& {Cooksey}}]{Prochaska:2011}
{Prochaska} J.~X., {Weiner} B., {Chen} H.-W., {Mulchaey} J., {Cooksey} K.,
  2011, ApJ, 740, 91

\bibitem[{{Reyes} {et~al.}(2012){Reyes}, {Mandelbaum}, {Gunn}, {Nakajima},
  {Seljak}, \& {Hirata}}]{Reyes:2012}
{Reyes} R., {Mandelbaum} R., {Gunn} J.~E., {Nakajima} R., {Seljak} U., {Hirata}
  C.~M., 2012, MNRAS, 425, 2610

\bibitem[{{Rudie} {et~al.}(2013){Rudie}, {Steidel}, {Shapley}, \&
  {Pettini}}]{Rudie:2013}
{Rudie} G.~C., {Steidel} C.~C., {Shapley} A.~E., {Pettini} M., 2013, ApJ, 769,
  146

\bibitem[{{Savage} {et~al.}(2014){Savage}, {Kim}, {Wakker}, {Keeney}, {Shull},
  {Stocke}, \& {Green}}]{Savage:2014}
{Savage} B.~D., {Kim} T.-S., {Wakker} B.~P., {Keeney} B., {Shull} J.~M.,
  {Stocke} J.~T., {Green} J.~C., 2014, ApJS, 212, 8

\bibitem[{{Schlegel} {et~al.}(1998){Schlegel}, {Finkbeiner}, \&
  {Davis}}]{Schlegel:1998}
{Schlegel} D.~J., {Finkbeiner} D.~P., {Davis} M., 1998, ApJ, 500, 525

\bibitem[{{Shectman} \& {Johns}(2003)}]{Shectman:2003}
{Shectman} S.~A., {Johns} M., 2003, in The International Society for Optical
  Engineering Proceedings, {Oschmann} J.~M., {Stepp} L.~M., eds., Vol. 4837,
  pp. 910--918

\bibitem[{{Shen} {et~al.}(2013){Shen}, {Madau}, {Guedes}, {Mayer}, {Prochaska},
  \& {Wadsley}}]{Shen:2013}
{Shen} S., {Madau} P., {Guedes} J., {Mayer} L., {Prochaska} J.~X., {Wadsley}
  J., 2013, ApJ, 765, 89

\bibitem[{{Stocke} {et~al.}(2013){Stocke}, {Keeney}, {Danforth}, {Shull},
  {Froning}, , \& {\etal}}]{Stocke:2013}
{Stocke} J.~T., {Keeney} B.~A., {Danforth} C.~W., {Shull} J.~M., {Froning}
  C.~S., , {\etal}, 2013, ApJ, 763, 148

\bibitem[{{Stocke} {et~al.}(2014){Stocke}, {Keeney}, {Danforth}, {Syphers},
  {Yamamoto}, {Shull}, {Green}, {Froning}, {Savage}, {Wakker}, {Kim},
  {Ryan-Weber}, \& {Kacprzak}}]{Stocke:2014}
{Stocke} J.~T., {Keeney} B.~A., {Danforth} C.~W., {Syphers} D., {Yamamoto} H.,
  {Shull} J.~M., {Green} J.~C., {Froning} C., {Savage} B.~D., {Wakker} B.,
  {Kim} T.-S., {Ryan-Weber} E.~V., {Kacprzak} G.~G., 2014, ApJ, 791, 128

\bibitem[{{Tejos} {et~al.}(2014){Tejos}, {Morris}, {Finn}, {Crighton},
  {Bechtold}, {Jannuzi}, {Schaye}, {Theuns}, {Altay}, {Le F{\`e}vre},
  {Ryan-Weber}, \& {Dav{\'e}}}]{Tejos:2014}
{Tejos} N., {Morris} S.~L., {Finn} C.~W., {Crighton} N.~H.~M., {Bechtold} J.,
  {Jannuzi} B.~T., {Schaye} J., {Theuns} T., {Altay} G., {Le F{\`e}vre} O.,
  {Ryan-Weber} E., {Dav{\'e}} R., 2014, MNRAS, 437, 2017

\bibitem[{{Tripp} {et~al.}(1998){Tripp}, {Lu}, \& {Savage}}]{Tripp:1998}
{Tripp} T.~M., {Lu} L., {Savage} B.~D., 1998, ApJ, 508, 200

\bibitem[{{Tumlinson} {et~al.}(2013){Tumlinson}, {Thom}, {Werk}, {Prochaska},
  {Tripp}, , \& {\etal}}]{Tumlinson:2013}
{Tumlinson} J., {Thom} C., {Werk} J.~K., {Prochaska} J.~X., {Tripp} T.~M., ,
  {\etal}, 2013, ApJ, 777, 59

\bibitem[{{Tumlinson} {et~al.}(2011){Tumlinson}, {Thom}, {Werk}, {Prochaska},
  {Tripp}, \& {\etal}}]{Tumlinson:2011}
{Tumlinson} J., {Thom} C., {Werk} J.~K., {Prochaska} J.~X., {Tripp} T.~M.,
  {\etal}, 2011, Science, 334, 948

\bibitem[{{Wakker} \& {Savage}(2009)}]{Wakker:2009}
{Wakker} B.~P., {Savage} B.~D., 2009, ApJS, 182, 378

\bibitem[{{Wang}(1993)}]{Wang:1993}
{Wang} B., 1993, ApJ, 415, 174

\bibitem[{{Werk} {et~al.}(2013){Werk}, {Prochaska}, {Thom}, {Tumlinson},
  {Tripp}, \& {\etal}}]{Werk:2013}
{Werk} J.~K., {Prochaska} J.~X., {Thom} C., {Tumlinson} J., {Tripp} T.~M.,
  {\etal}, 2013, ApJS, 204, 17

\bibitem[{{Werk} {et~al.}(2012){Werk}, {Prochaska}, {Thom}, {Tumlinson},
  {Tripp}, {O'Meara}, \& {Meiring}}]{Werk:2012}
{Werk} J.~K., {Prochaska} J.~X., {Thom} C., {Tumlinson} J., {Tripp} T.~M.,
  {O'Meara} J.~M., {Meiring} J.~D., 2012, ApJS, 198, 3

\bibitem[{{Wild} {et~al.}(2008){Wild}, {Kauffmann}, {White}, {York}, {Lehnert},
  {Heckman}, {Hall}, {Khare}, {Lundgren}, {Schneider}, \& {vanden
  Berk}}]{Wild:2008}
{Wild} V., {Kauffmann} G., {White} S., {York} D., {Lehnert} M., {Heckman} T.,
  {Hall} P.~B., {Khare} P., {Lundgren} B., {Schneider} D.~P., {vanden Berk} D.,
  2008, MNRAS, 388, 227

\bibitem[{{Yoon} {et~al.}(2012){Yoon}, {Putman}, {Thom}, {Chen}, \&
  {Bryan}}]{Yoon:2012}
{Yoon} J.~H., {Putman} M.~E., {Thom} C., {Chen} H.-W., {Bryan} G.~L., 2012,
  ApJ, 754, 84

\bibitem[{{York} {et~al.}(2000){York}, {Adelman}, {Anderson}, \&
  {\etal}}]{York:2000}
{York} D.~G., {Adelman} J., {Anderson} Jr. J.~E., {\etal}, 2000, AJ, 120, 1579

\end{thebibliography}
}

\bsp \label{lastpage}

\end{document}